\documentclass{egpubl}
\usepackage{VMV2024}
 
\WsPaper           

\usepackage[T1]{fontenc}
\usepackage{dfadobe}  

\usepackage{cite}  
\BibtexOrBiblatex
\electronicVersion
\PrintedOrElectronic
\ifpdf \usepackage[pdftex]{graphicx} \pdfcompresslevel=9
\else \usepackage[dvips]{graphicx} \fi

\usepackage{egweblnk} 

\title[Enhancing Uncertainty Communication in Time Series Predictions: Insights and Recommendations]%
      {Enhancing Uncertainty Communication in Time Series Predictions: Insights and Recommendations} 

\author[Karagappa et al.]
{\parbox{\textwidth}{\centering A. Karagappa$^{1}$,
        P. Kaur Betz$^{1}$,
        J. Gilg$^{1}$,
        M. Zeumer$^{1}$,
        A. Gerndt$^{1}$ and
        B. Preim$^{2}$
        }
        \\
{\parbox{\textwidth}{\centering $^1$Institute for Software Technology, DLR, Braunschweig, Germany\\
         $^2$Faculty of Computer Science, University of Magdeburg, Magdeburg, Germany
       }
}
}

\usepackage{caption}
\usepackage{xcolor}
\usepackage{amsmath}
\begin{document}


\maketitle
\begin{abstract}
As the world increasingly relies on mathematical models for forecasts in different areas, effective communication of uncertainty in time series predictions is important for informed decision making. This study explores how users estimate probabilistic uncertainty in time series predictions under different variants of line charts depicting uncertainty. It examines the role of individual characteristics and the influence of user-reported metrics on uncertainty estimations. By addressing these aspects, this paper aims to enhance the understanding of uncertainty visualization and for improving communication in time series forecast visualizations and the design of prediction data dashboards. \\

\ccsdesc[500]{Human-centered computing~Visualization theory, concepts and paradigms}
\ccsdesc[500]{Human-centered computing~Empirical studies in visualization}
\ccsdesc[500]{Human-centered computing~Visualization design and evaluation methods}

\printccsdesc   
\end{abstract}  

\section{Introduction} \label{sec:intoduction}

\begin{figure*}[t]
    \centering
    \includegraphics[scale = 0.25]{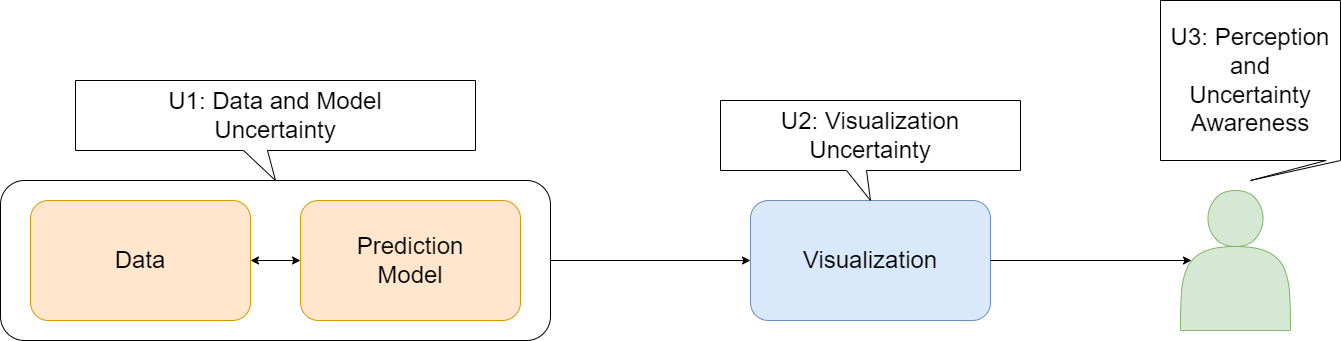}
    \caption{A simplified version of the knowledge generation model \cite{Sacha2016_US}}
    \label{fig:uncertainty_pipeline}
\end{figure*}

In time series forecasts, a model predicts the future values of a variable based on its past data as well as other contributing factors, using statistical or machine learning techniques. It finds application in several areas including weather forecasting, predicting growth indicators (e.g. GDP) and predicting daily cases of infectious diseases (e.g. COVID-19). These forecasts are always accompanied by uncertainty. The uncertainty can arise in many different stages of the visualization pipeline, see Figure \ref{fig:uncertainty_pipeline} \cite{Sacha2016_US}. In the figure, there are three types of uncertainty depicted:

(i) Data and model uncertainty (\textbf{U1}): This uncertainty arises from factors like data collection, preprocessing, the prediction model, and parameterization.

(ii) Visualization uncertainty (\textbf{U2}): This uncertainty stems from how uncertainty is visually encoded and how our brains naturally perceive it.

(iii) Perception and uncertainty awareness (\textbf{U3}): This uncertainty relates to differences between users due to individual characteristics between them. 

\par Often the visualization authors do not have control over the data and model uncertainty, therefore, the goal is to clearly communicate this while minimizing the visualization uncertainty and aiding the perception and uncertainty awareness in the user. 

\begin{figure}[h!]
    \centering
    \includegraphics[scale=0.35]{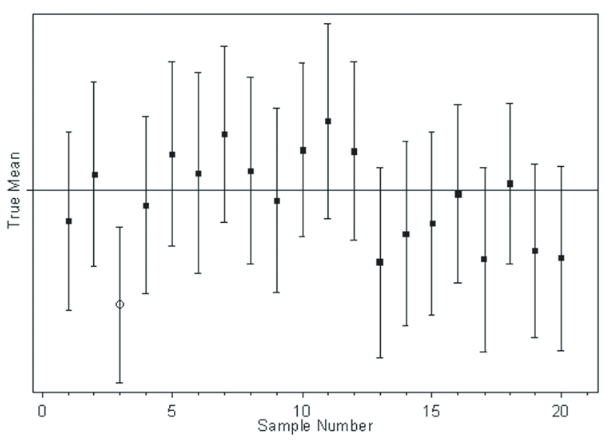}
    \caption{For 20 independent samples from the population, nineteen of their 95\% confidence intervals (marked by black squares) contain the true mean, while one (marked by a hollow circle) does not \cite{tantan_ci}.}
    \vspace{-10mm}
    \label{fig:what_is_a_Ci}
\end{figure}

\par The uncertainty is often expressed in terms of a credible or a confidence interval. In a literature review on the reliability of predictions on COVID-19, the reliability of cumulative cases was found to be reported using either confidence intervals (CI) or credible intervals (CrI) \cite{GNANVI2021258}. While both the CI and CrI are measures of uncertainty, they differ in their nature. A 95\% CrI suggests there is a 95\% probability that the true value lies within the predicted interval given the distribution of the observed data. This interpretation is often applied to a confidence interval, but is incorrect \cite{Hoekstra2014-confidenceinterval}. A 95\% CI implies that when a study is conducted multiple times, each time with different samples from the same population, each sample will produce its own 95\% CI. We expect 95\% of these CIs to contain the true population mean, and this expectation holds with 100\% certainty, see Fig \ref{fig:what_is_a_Ci} \cite{tantan_ci}.

\par Given the wide range of applications, naturally, the users are diverse. It is essential that the uncertainty accompanying time series visualizations is understood by the diverse users and considered for decision making. To achieve this, we must ensure that: 1. the graphical representation of uncertainty clearly communicates its nature and value, and 2. the complete information required by the user for their decision making tasks is made available alongside this graphical representation.
\par Graphically, the uncertainty in time-series forecasts is often represented with a confidence band in a line chart (see Figure \ref{fig:esid}), or some variation of it. Previous research has shown that with some visualizations, users are more likely to adopt an interpretation of uncertainty that closely resembles a normal distribution. It is important to understand which individual properties of a user contributes to their interpretation of uncertainty and their accuracy in estimating it from a visualization. This should allow us to design the uncertainty visualization so that disparities between user perceptions of uncertainty may be minimized. Additionally, it is unclear whether the nature and complexity of the uncertainty are communicated well enough, contextually, for a user to include the uncertainty estimation in their decision-making process.
\par In order to answer these questions, we conducted two user studies, with observed variables on individual characteristics, task performance, and user-reported metrics. The results of these studies have found evidence in support of five major considerations in developing visualizations for prediction charts: 
\begin{itemize}
    \item Developing and utilizing standard uncertainty terminology and visualization techniques
    \item Meeting the different information needs of users
    \item Neutralizing the effects of numeracy
    \item Emphasis on clutter reduction and aesthetics in design
    \item The inclusion of interactive techniques for increased comprehensibility
\end{itemize}

\section{Motivation}

\begin{figure*}
    \centering
    \includegraphics[scale = 0.34]{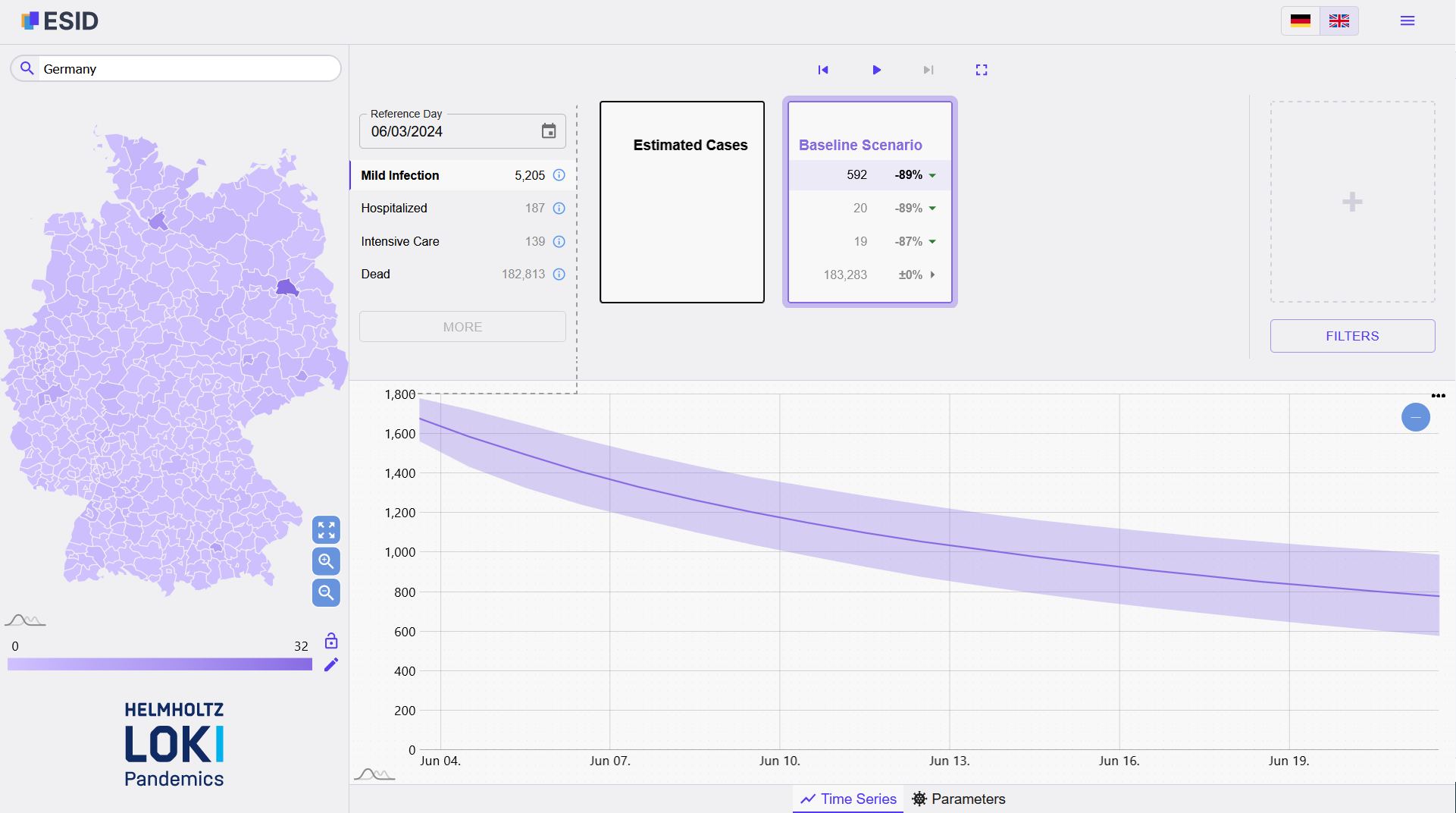}
    \setlength{\belowcaptionskip}{-10pt}
    \caption{A snapshot of the Epidemiological Scenarios for Infectious Diseases (ESID) application, a visual analytics application developed to show the
results of epidemiological simulations. On the left hand side is a map of the German counties with a customizable heat legend. At the top section is a purple scenario card representing one simulation scenario. Left of the scenario cards is a list of infection states or aggregations of infection states, like infected, hospitalized, and dead. The bottom portion contains a line chart comparing the scenarios over a predefined timeline. \cite{betz2023esid}.}
    \label{fig:esid}
\end{figure*}

In 2020, we saw the rise of many Public Health Dashboards that displayed the predicted daily cases, hospitalizations and even deaths due to COVID-19. The COVID-19 Projections dashboard by the Institute for Health Metrics and Evaluation (IHME) \cite{ihme_covid19}, was likely the first dashboard available for providing forecasts related to the COVID-19 pandemic. WHO/Europe stated that since the start of the COVID-19 pandemic, it increasingly referenced IHME forecasts in providing guidance and recommendations to countries in the WHO European region with respect to containing the pandemic \cite{who_ihme_agreement}. 
\par Undeniably, these visualizations played a significant role in public safety and well-being. Simultaneously, there were concerns over the validity of the predictions and their usefulness to policymakers. One particular concern was whether the graphical representation of uncertainty in line charts is effectively conducive to understanding uncertainty in predicted peak daily values, with instances of the uncertainty bounds, and consequently the projections, being interpreted incorrectly in both formal and social media \cite{jewell_concerns_over_covid_projections}. 
\par This concern is heightened by reports from other national-level COVID-19 dashboard development teams, indicating challenges in presenting data clearly and understandably \cite{barbazza_33_dashboards_study}. The challenges faced by dashboard development teams were likely exacerbated by the rapidly evolving nature of the COVID-19 virus and time constraints in the design and development process. Additionally, authors generally omit uncertainty from their visualizations due to lack of canonical forms of uncertainty visualizations and expected difficulty for users in interpreting them \cite{hullman_authors_dont_viz_uncertainty}. 

In their examination of Epidemiological Scenarios for Infectious Diseases (ESID), a visual analytics tool aimed at projecting simulated developments of infectious disease spread (see Figure \ref{fig:esid}), Betz et al. discovered issues of misinterpretation and misuse arising from inadequate treatment and communication of uncertainty \cite{betz2023esid}. As a result, we find the need to establish guidelines that help visualization authors effectively communicate uncertainty based on empirical findings. 

\section{Related Work}
For representing uncertainty, Olston and Mackinlay highlighted the necessity for distinct visual representations of bounded and statistical uncertainty, addressing the potential misinterpretation that led to the development of the confidence interval band\cite{olsten_ambiguity}. Subsequent studies revealed variations in the interpretation of confidence bands and its variants, with some aligning more closely with a credible interval\cite{STak2014_UserStudy}. Although the results provide valuable insights, the recent misinterpretation of credible intervals underscores the need for a study where uncertainty is precisely defined, and suitable representations are identified to ensure comprehensive information for user trust. An individual's numeracy influences their uncertainty estimation\cite{STak2014_UserStudy, AToet2019}, thus there is a desire to explore whether numeracy is impacted by other factors. This exploration aims to inform the design of visualizations that facilitate training or support interpretation for a broad user base. 

Trust in uncertainty visualizations and visual processing has been a subject of study\cite{elhamdadi2022}. Yet another crucial aspect is whether users feel assured in their understanding of the given information\cite{Yeung2012-gu}. Consequently, we investigate to determine if such assurance can be influenced by specific characteristics of the visualization. While visualization authors may face limitations in the available information \cite{hullman_authors_dont_viz_uncertainty}, the goal is to assist them in designing dashboards that consider factors within their control.

Lastly, we have realised a gap in the literature concerning evaluations of whether prediction dashboards adequately present the information needed to make decisions based on uncertainty. Our studies explored these gaps and sought to identify their potential sources. 
\section{Method}
Our work focuses on two sets of factors that inform the design process of uncertainty visualizations: those contributing to visual uncertainty (U2 in Figure \ref{fig:uncertainty_pipeline}) and those affecting uncertainty perception of individuals (U3 in Figure \ref{fig:uncertainty_pipeline}). The factors considered include:
\begin{itemize}
    \item Visual uncertainty factors:
    \begin{itemize}
        \item Task performance under different techniques
        \item User-reported metrics of visualization qualities, such as clutter and aesthetic.
    \end{itemize}

    \item Uncertainty perception factors:
    \begin{itemize}
        \item Individual characteristics, including numeracy, field of work, and familiarity with line charts.
        \item Information needs
        \item User-reported metrics of user performance, such as success and difficulty.
    \end{itemize}
\end{itemize}

\noindent To summarize, the questions we want to answer are:
\begin{itemize}
    \item \noindent \textbf{Task Performance:} What is the impact of different visualization techniques on the participants’ uncertainty estimation?
    \item \textbf{Information Needs:} Does the information provided by the uncertainty predictions visualizations meet the needs of the users? 
    \item \textbf{User-Reported Metrics:} How do the effects of varying visualization techniques (like clutter and aesthetics) influence users' evaluations of task difficulty, and their perceived level of success in task performance? 
    \item \noindent \textbf{Individual Characteristics:} Do individual characteristics shared among target user groups, such as area of study, frequency of visualization use, the highest degree correlate with numeracy, and subsequently task performance? 
\end{itemize}

\noindent To answer our research questions, we conducted one user study with a diverse user base and a second user study with the target user group. The target user group consists of participants from local health authorities, medicine and neuroscience. These individuals typically work with medical images, epidemiological data, and uncertainties within their respective fields. They also carry some degree of job responsibility in being able to understand and interpret epidemiological predictions. In the design of public health dashboards, they are usually among the main stakeholders \cite{betz2023esid}, and thus results from them are deemed important in the visualization design.  

\subsection{Stimuli}
\begin{figure}[t]
    \centering
    \includegraphics[width=8cm]{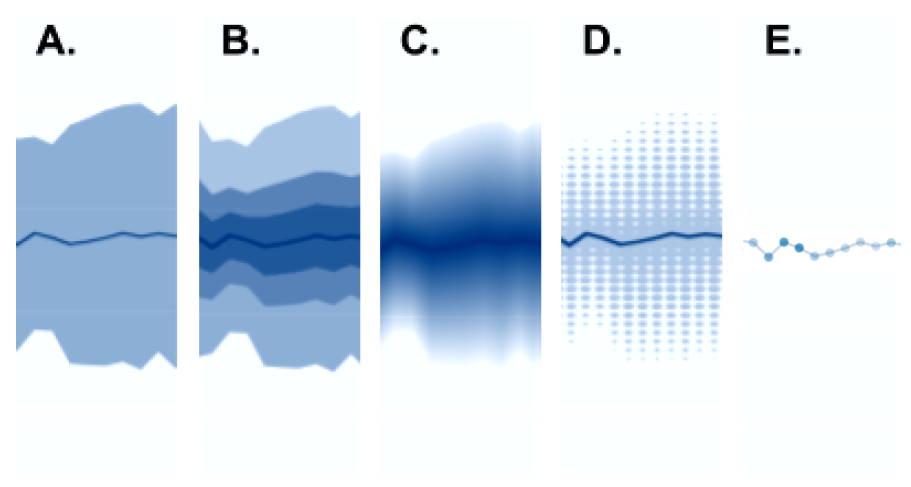}
    \caption{Uncertainty represented as A. Confidence Band, B. Overlapping Bands, C. Blur, D. Circular Glyphs and E. Colored Markers}
    \label{fig:visualisation_types}
\end{figure}

In the initial user study, we inspected five uncertainty visualizations, as shown in Figure \ref{fig:visualisation_types}. The techniques chosen are A. Confidence Band \cite{olsten_ambiguity}, B. Overlapping Bands of varying credible intervals, C. Band represented by decreasing colour saturation, D. Band represented by decreasing area of circular glyph and E. Colored Markers of varying colour saturation \cite{sanyal_marker}.

The user studies consisted of three parts; collection of participants' data, a numeracy task, and visualization tasks. 
\subsection{Participants Data}
The collection of individual information included participants’ highest degree, field of study and the frequency with which they interact with visualizations (see Q.P1 - Q.P3, Supplementary Material).
\subsection{Numeracy Task}
\par To measure the users' numeracy, we adopted three questions from the J. Weller et al. numeracy scale whose difficulties span over easy, intermediate and difficult (see Q.N1 - Q.N3, Supplementary Material) \cite{JWeller2013}. For the final study, we included two additional questions to assess their comprehension of a normal distribution (see Q.N4 a and b, Supplementary Material).
\subsection{Visualization Task}
\par The visualization task was divided into two subtasks. The first subtask featured three questions concerning probability estimation (see Q.T1, Q.T4 and Q.T7, Supplementary Material). The participants were tasked with assessing the likelihood with which the predicted value falls within a certain range. This is done in a decision-making context, although no decision is asked to be made \cite{JHullman2019_EvaluationSurvey, SCastro2022_UserStudy}. The uncertainty estimations are elicited in the form of a range slider with the minimum and maximum marked at 0 and 100, respectively. Following each probability estimation, users were asked to declare whether the information provided was enough to solve the task (see Q.T2, Q.T5 and Q.T8, Supplementary Material). Only in the final study, if they required more information, they were prompted to describe what particular information they needed (see Q.T3, Q.T6 and Q.T9, Supplementary Material). 
\par The second subtask focused on gathering user-reported metrics. Two questions pertained to the task (see Q.T10a and b, Supplementary Material) and two questions pertained to the characteristics of the visualizations (see Q.T10 c and d, Supplementary Material). Questions Q.T10 a and b seek to evaluate the users' perceived difficulty of the task and their perceived success in the task, taken from the NASA-TLX questionnaire \cite{NASATLX}. Questions Q.T10 c and d seek to evaluate the users' perceived clutter in and perceived aesthetic of the visualization technique. This is essential to understanding what aspects of a visualization an author must pay attention to in order to improve the communication of uncertainty \cite{Sacha2016_US}.  The four user-reported metrics were measured on a five point Likert scale.

\section{Results}
The evaluation of the results was performed using the R programming language and associated statistical tools and libraries. 

\subsection{Measures}

\par In our studies, all participants were exposed to all visualization techniques, and the resulting data was not normally distributed. Therefore, to determine if there were significant differences in uncertainty estimations across these techniques, we conducted a Friedman test using the \texttt{stats::friedman.test} function. Following this, we performed a post hoc Nemenyi test using the \texttt{PMCMRplus::frdAllPairsNemenyiTest} function to identify significant pairwise differences between the various uncertainty visualization methods.

The performance metric used to validate the uncertainty estimation was the normalized sum of all three uncertainty estimations under a visualization technique. The normalization was performed by calculating the range of the sum of the uncertainty estimations across all participants, then applying the following formula:

\[
\text{Performance} = \frac{\max(U_1 + U_2 + U_3) - (U_1 + U_2 + U_3)}{\max(U_1 + U_2 + U_3) - \min(U_1 + U_2 + U_3)}
\]

where \( U_1 \), \( U_2 \), and \( U_3 \) represent the three uncertainty estimations under each visualization for each participant. \( U_N \) (where \( N = 1, 2, 3 \)) is calculated as the absolute difference between the user estimate and the true value. This was chosen as the performance measure as it considers the overall uncertainty by aggregating estimations from different ranges of uncertainty variation across techniques. Furthermore, normalization enhances interpretability by scaling the metric to a common range, facilitating clearer insights into relative performance.

Lastly, Spearman Rho is used to find correlation between ordinal-ordinal and continuous-ordinal values, utilizing the \texttt{stats::cor} function. Ordinal encoding is employed where necessary. This correlation can be seen in Figure \ref{fig:combined_corr_matrix}, where only the correlations with a \textit{p} value below 0.05 are visualized. The \textit{p} values are calculated using the Spearman method from the \texttt{rstatix::cor\_pmat} function. The correlation values are interpreted according to Table \ref{tab:spearman_correlation} \cite{spearman_rho_interpretation_dancey_reidy}.

\begin{table}[htbp]
\centering
\begin{tabular}{|c|l|}
\hline
\textbf{Spearman's Rho} & \textbf{Interpretation}               \\ \hline
0.01 - 0.19              & None or very weak relationship       \\ \hline
0.20 - 0.29              & Weak relationship                    \\ \hline
0.30 - 0.39              & Moderate relationship                \\ \hline
0.40 - 0.69              & Strong relationship                  \\ \hline
$\geq$ 0.70              & Very strong relationship             \\ \hline
\end{tabular}
\caption{Interpretation of Spearman's Rho correlation \cite{spearman_rho_interpretation_dancey_reidy}}
\label{tab:spearman_correlation}
\end{table}

Points are awarded in the numeracy task based on question difficulty: 1 point for the easiest question, 2 points for the next, and so on. The lowest possible score is 0. In the initial study with three questions, participants could score up to 6 points, with a mean possible score of 3. In the second study with five questions, the maximum score was 15 points, with a mean possible score of 7.5.

\subsection{Initial User Study With a Diverse Audience}
\par
\begin{figure}
    \centering
    \includegraphics[width=0.33\textwidth]{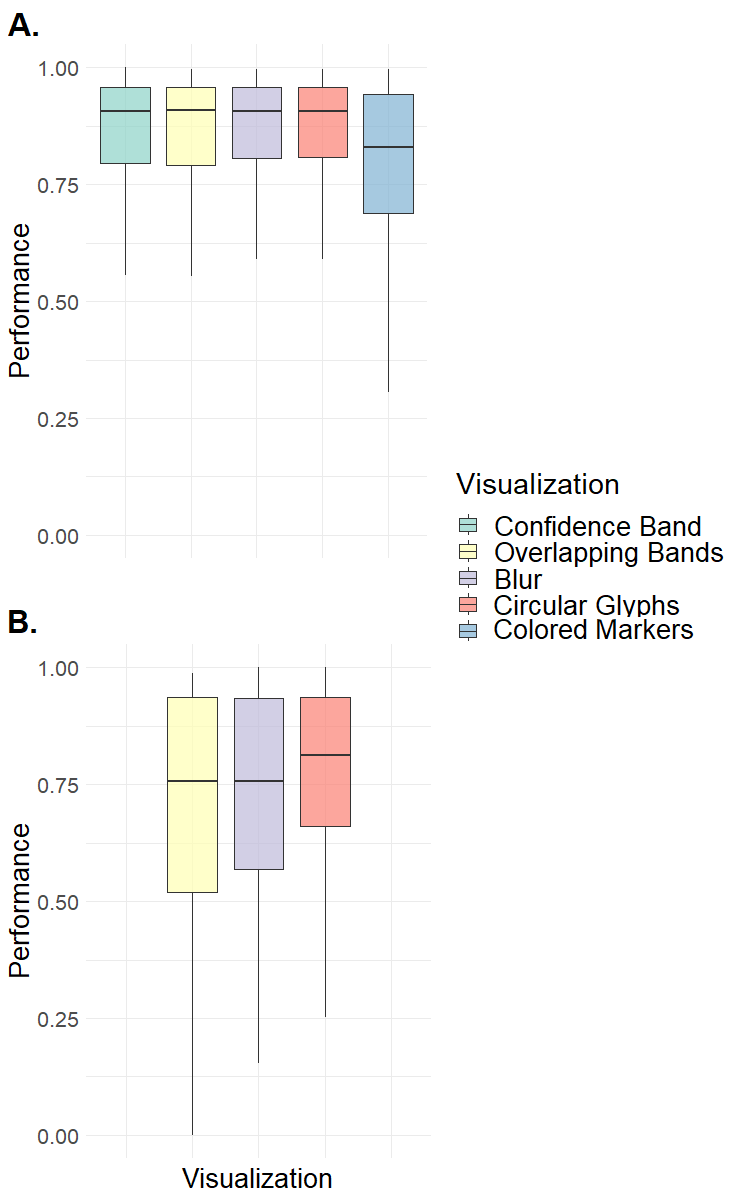}
    \setlength{\belowcaptionskip}{-13pt}
    \caption{Boxplots showing the performance of visualization techniques. \textbf{A:} The five visualization techniques from the initial study ($p < 0.001$). The performance under \textit{Colored Marker} is significantly worse than all the other techniques. \textbf{B:} The three visualization techniques from the second study ($p \leq 0.05$). \textit{Colored Marker} is not considered for the second study due to its poor performance in the initial study. \textit{Confidence Band} is also excluded as it did not show a significant difference compared to other techniques and does not qualitatively depict the nature of a credible interval. In the second study, \textit{Circular Glyphs} shows a significant difference with \textit{Blur} ($p = 0.058$) but no other significant differences were observed. \textit{Circular Glyphs} demonstrates higher median performance and smaller variability.}
    \label{fig:vis_perf_prelim}
\end{figure}

 The initial user study involved 94 participants from diverse fields, ranging from high school to individuals holding a PhD or advanced degrees. The Friedman test shows significant ($p < 0.001$) differences between the performance of users with different visualization techniques. A post hoc Nemenyi test finds significant ($p \leq 0.001$) differences between \textit{Colored Marker} and all other techniques. As shown in Figure \ref{fig:vis_perf_prelim}A, the \textit{Colored Marker} visualization technique performed the worst in the initial user study. This could be attributed to the lack of area representation in \textit{Colored Marker} that might have allowed users to build heuristics to estimate uncertainty in the other techniques. Therefore, in the second user study, we do not consider the \textit{Colored Markers} technique owing to its bad performance. We also omitted \textit{Confidence Band} as it did not show significantly better performance than other techniques and also does not qualitatively depict the nature of a credible interval. 

\begin{figure*}[t]
  \centering
    \includegraphics[scale=0.45]{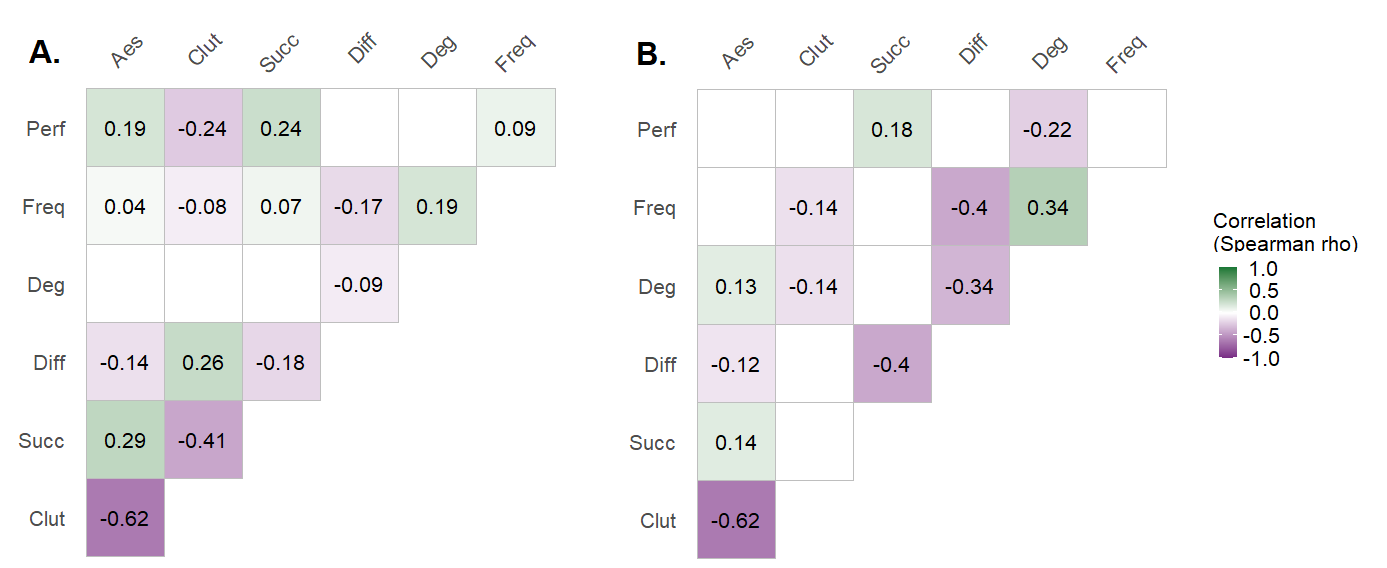}
    \setlength{\belowcaptionskip}{-12pt}
    \caption{Correlation matrix of performance, individual differences, and user-reported metrics in \textbf{A}. the initial study and \textbf{B}. the second study. Abbreviations: Perf (Performance of user in uncertainty estimation tasks, with range from Worst to Best), Freq (Frequency with which the user interacts with visualizations, with range from Rarely to Once a week), Deg (Highest Degree held by the user, with range from High School to PhD or Higher), Diff (Perceived difficulty of task given a visualization technique, with range from Not hard at all to Very hard), Succ (Perceived success of task given a visualization technique, with range from Poor to Good), Clut (Clutter of a visualization technique, with range from Uncluttered to Cluttered), Aes (Aesthetic of a visualization technique, with range from Bad to Good).}
    \label{fig:combined_corr_matrix}
\end{figure*}

Additionally, in the initial study, we also observe that perceived aesthetics and perceived success exhibit a strong negative correlation with perceived clutter (see Figure \ref{fig:combined_corr_matrix}A). Visual information processing in novices have shown to be negatively impacted by clutter in financial visualizations, in comparison to experts\cite{display_clutter}. Additionally, aesthetic visualizations have previously shown to display a higher level of user patience, resulting in lower task abandonment and erroneous response \cite{aesthetic}.

\subsection{Second User Study With the Target User Group} 
To know the information needs of the specific group, the next study was conducted with 31 participants. In this study, we compared \textit{Blur}, \textit{Overlapping Bands} and \textit{Circular Glyphs} (C, B and D in reference to Figure \ref{fig:visualisation_types}). In the following, we  present the results according to the objectives defined earlier.

\noindent  \textbf{Task performance:} The Friedman test shows significant ($p < 0.05$) differences between the performance of users with different visualization technique. A post hoc Nemenyi test reveals a trend towards significant differences (p = 0.058) between  \textit{Circular Glyphs} and  \textit{Blur}.  \textit{Circular Glyphs} demonstrate a higher median performance and lower variability compared to  \textit{Blur} and  \textit{Overlapping Bands} (see Figure \ref{fig:vis_perf_prelim}B). It is challenging to understand why, because there was no observed correlation between the visualization technique and the factors that influence usability. Earlier empirical findings demonstrate that, in quantitative perceptual tasks, the use of area outperforms colour hue\cite{cleaveland_and_gill_graphical_ranking}. This observation may offer insights into why \textit{Circular Glyphs} is more effective than \textit{Blur}. Furthermore, minor inaccuracies in estimating the size of an area result in only minor misperceptions of the associated quantitative value being encoded \cite{mackinlay_automation_graphical}.

\noindent \textbf{Information needs:} In approximately a third of the uncertainty estimation task instances, participants reported that they did not have the necessary information to make the required estimate. The information needed by participants can be divided largely into two groups: Statistical Information and Model Information. Statistical Information refers to the information needed to make the correct association between a credible interval and likelihood, while Model Information refers to the information regarding parameters chosen and model used in the prediction. 

\noindent \textbf{User-reported metrics:} Similar to the findings of the first study, we observe a strong correlation between perceived clutter and aesthetic. However, perceived success does not correlate with clutter; instead, it correlates with perceived difficulty. This perceived difficulty, in turn, correlates with individual characteristics, such as a user's degree and frequency of interaction with line charts.

\noindent \textbf{Individual characteristics:} Users' frequency of interaction with line charts has a weak correlation with their numeracy (Rho = 0.27). Higher numerate users (numeracy $>$ 7.5) exhibit lower errors across all visualizations combined, with a narrow error range. In contrast, those with low numeracy show higher and more variable errors.

\begin{figure}
    \centering
    \includegraphics[scale=0.36]{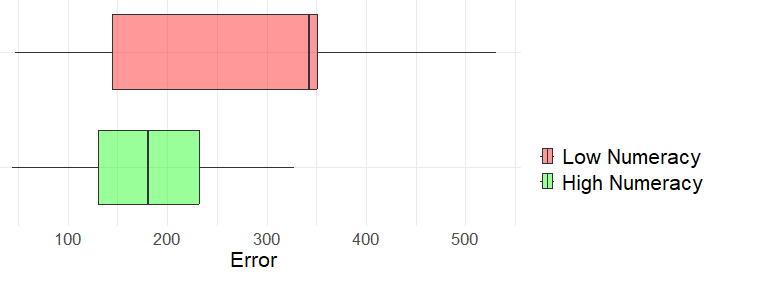}
    \caption{Boxplots comparing total user errors across all visualization techniques combined between higher numerate (numeracy $>$ 7.5) and lower numerate users.}
    \label{fig:enter-label}
\end{figure}



\section{Guidelines}
\par Given the results of the user study and the relationship between the measured metrics, we have formulated four guidelines for uncertainty visualization designers.
\subsection{Standardizing Uncertainty Terminology and Visualization Techniques}\label{sec:standardize}
    \par There exists a discrepancy in the terminology used to describe uncertainty of predictions in the present literature, including but not limited to: Credible Interval, Prediction Interval, Simulation Percentile, and Uncertainty Interval. While the choice of terminology used for uncertainty ranges depends on the specific modeling approach and uncertainty quantification, both researchers and users of visualizations can benefit from standardizing the terminology. When illustrating uncertainty ranges visually, it is important to balance clarity and completeness. Whether it is a precise interval or a probabilistic distribution, the graphical representation and terminology should accurately reflect this.

\subsection{Meeting the Informational Needs of a Diverse Audience}
For users to make an informed decision, it is advisable to include the following information:
\begin{itemize}
    \item \textbf{Statistical Information}
    \par  In the second user study, participants either explicitly stated or hinted towards needing more statistical information that would help them link the terms \textit{likelihood} and \textit{uncertainty interval}. A participant suggested a calculation tool to help make this connection. The uncertainty showed was credible intervals, but some participants expressed confusion regarding the distinction between confidence intervals and credible intervals. 
\item \textbf{Model Information} 
\par Participants in the second user study also stated that the information provided about the parameters in the prediction model and its \textit{historic accuracy} was necessary to adequately make an estimation. Providing access to previous forecasts with uncertainty in conjunction with truth data could be used to show historic accuracy. 
\end{itemize}

Tool-tips or hover-over explanations, annotations, and narrative approaches should be considered to fill these gaps in knowledge.

\subsection{Equalizing Effects of Numeracy in Uncertainty Estimation}
    Minimizing the cognitive effort needed for making many inferences is beneficial to users with lower numeracy \cite{EPeters2007}. The role of uncertainty and the risk of excluding it in the decision-making process should be explicitly displayed as people tend to underweight probable outcomes compared to certain ones\cite{paper23}. Consider hiding visual objects that do not contain relevant information when presenting uncertainty. Removing non-quality information will improve comprehension for those with lower numeracy skills. Additionally, use consistent and comprehensible formats. For instance, simplify uncertainty ranges to ensure they are easily understandable and manageable. 

\subsection{Clutter Reduction and Aesthetic Design Influence on Uncertainty Estimation}
    Incorporating the range of uncertainty intervals and avoiding the overlap of uncertainties should be integral to the design process. If not carefully managed, improper handling of these elements can lead to axis realignment and overwhelming information, thereby compromising the effectiveness of the visualization. Decluttering graphs involves removing backgrounds, chart borders, gridlines, and unnecessary axis lines, adding white space between major elements and replacing legends with direct labels \cite{paperA}. In a study conducted, grayscale visualizations of ensemble predictions were more trusted than colour-coded visualizations \cite{padilla_mfv}. The authors initially predicted that adding colour may add complexity and reduce clarity, whose effect would increase with the number of forecasts but found it to be consistently true irrespective of the number of forecasts. 
\subsection{Increase Comprehensibility Through Interaction}
\par Provide facilities for better comprehension of uncertainty visualizations. For instance, a visualization may incorporate two movable horizontal lines that allow users to render a calculated probability of the predicted value of each displayed model falling within the selected range defined by these lines. Rather than policy makers having to examine the given information and make educated guesses about the uncertainties associated with various intervention strategies to identify the most suitable one, they can adopt a reverse approach based on their existing data. They can start by stating, for example, that they currently have X hospital beds available and the goal is to limit the number of infections to Y to accommodate this capacity. Then, they can evaluate various predictions resulting from varying intervention plans and select the one that best aligns with these predefined requirements. In essence, it simplifies decision making by working backward from their existing constraints and goals.
\section{Conclusion and Future Work}

In this paper, we presented the results of our two user studies which we conducted to understand the factors influencing visual uncertainty and individual uncertainty perception. The initial study involved 94 participants from diverse backgrounds, assessing five different uncertainty visualization techniques. The results indicated that   \textit{Circular Glyphs} and  \textit{Blur} showed the best performance, while the Colored Markers technique performed the worst. We found that higher numeracy and frequent interaction with visualizations positively correlated with better task performance and perceived success. Additionally, increased perceived clutter and decreased aesthetics negatively impacted perceived difficulty and actual performance. The second user study targeted 31 participants from local health authorities, medicine, and neuroscience, focusing on the three best-performing techniques from the initial study:  \textit{Blur},  \textit{Overlapping Bands}, and  \textit{Circular Glyphs}. In this study,  \textit{Circular Glyphs} slightly outperformed the other techniques. Participants frequently reported needing more statistical and model information to make informed uncertainty estimations. Similar to the initial study, higher numeracy correlated with better performance, and clutter and aesthetics influenced perceived difficulty and success.

\par Our user studies, aligned with recent research on uncertainty visualization evaluation, revealed key insights for designing dashboards
with time series predictions. While we focused on epidemiological forecasts, our guidelines are not limited to a specific domain. Instead, they aim to aid decision making by providing necessary information with minimal cognitive effort. In 
summary, our findings offer practical design guidance to visualization authors to maximize the utility of the information they have access to but within the constraints they face.

\par Future research should explore interactive tools to enhance comprehension of uncertainty visualizations, particularly for users with varying levels of numeracy. Additionally, further studies could investigate the impact of standardized terminology and consistent exposure to uncertainty visualizations to improve overall interpretability and usability across diverse user groups.

\bibliographystyle{eg-alpha-doi} 
\bibliography{library}       



\end{document}